\documentclass[aps,pra,twocolumn,showpacs,preprintnumbers,amsmath,amssymb]{revtex4} 
\usepackage{graphicx}
\usepackage{graphics}
\usepackage{psfrag}
\usepackage{dcolumn}
\usepackage{bm}
\usepackage{subfigure}

\begin{document}
\bibliographystyle{prsty}

\title{Correlation dynamics of strongly-correlated bosons in time-dependent optical lattices}
\author{Karen Rodr\'{\i}guez, Arturo Arg\"uelles and Luis Santos}
\affiliation{Institut f\"ur Theoretische Physik, Leibniz Universit\"at Hannover, Appelstr. 2, D-30167, Hannover, Germany}

\date{\today}

\begin{abstract}
We analyze by means of Matrix-Product-State simulations the correlation dynamics of strongly-correlated superfluid Bose gases in one-dimensional time-dependent optical lattices. We show that, as for the 
case of abrupt quenches, a quasi-adiabatic modulation of the lattice is characterized by a 
relatively long transient regime for which quasi-local single-particle correlation functions have already converged to a new equilibrium, whereas long-range correlations and particularly the quasi-condensate fraction may still present a very significant dynamics well after the end of the lattice modification. 
We also address the issue of adiabaticity by considering the fidelity between the time-evolved state and the ground-state of the final lattice.
\end{abstract}


\maketitle

\section{Introduction}


Strongly-correlated atomic gases in optical lattices constitute one of the most active fields 
in the physics of cold gases. Nowadays spectacular progresses allow for an unprecedented 
degree of control of these systems, which permit the detailed analysis of many-body phenomena as e.g. the realization of the superfluid (SF) to Mott insulator (MI) transition in bosonic lattice gases~\cite{Greiner2002}, the 3D fermionic Mott insulator~\cite{Jordens2008,Schneider2008} or the Tonks  gas~\cite{Paredes2004}. 


The high tunability and long characteristic time scales of these systems offer an ideal scenario to investigate non-equilibrium dynamics in a way not available in traditional condensed matter systems. In particular, lattice hopping rates may be easily tuned by modulating the intensity of the lasers creating the optical lattice, and the interactions may be also modified in real time by means of Feshbach resonances and time-dependent magnetic fields. These changes may be 
produced fast enough to be considered as a sudden quench. These quenches have attracted a growing attention in recent years, 
in particular in what concerns the evolution of correlations and possible equilibration after a quench~\cite{Calabrese2006,Cazalilla2006,Rigol2007,Kollath2007,Manmana2007,Lauchli2008,Moeckel2008,Manmana2009}. Thermalization (or actually the absence of it) was recently studied in a milestone experiment performed in nearly integrable one-dimensional Bose gases by Kinoshita et al~\cite{Kinoshita2006}. 


On the contrary, if the modification of the system is very slow, 
much slowlier than the tunneling rate, one can in principle assume the evolution as adiabatic~\cite{Blakie2004,Rey2006,Pollet2008}. 
The issue of adiabaticity is however far from trivial, especially in 
low dimensional gapless systems, as recently discussed by Polkovnikov and Gritsev~\cite{Polkovnikov2008}. 
Interestingly, in the so-called non-adiabatic scenarios, the adiabatic limit cannot be reached 
no matter how slow the change is introduced. Although this result is strictly speaking only 
applicable in integrable harmonic systems, it was shown in Ref.~\cite{Polkovnikov2008} 
that this non-adiabatic scenario may be obtained by considering an initial 
non-interacting 1D or 2D Bose gas in a lattice under a slow increase of the interaction strength. 
The harmonic approximation remains accurate as long as $U_0/Jn_0\ll 1$, where (see below) $U_0$ 
characterizes the on-site interactions, $J$ is the hopping rate, and $n_0$ is the filling factor. 
In this regime the truncated Wigner approximation (TWA), plus an additional first-order quantum correction, allows for an accurate description of the correlation dynamics~\cite{Polkovnikov2008}.

In this paper we study the correlation dynamics of a superfluid Bose gas in a 1D lattice 
during and after a modification of the lattice depth. Contrary to the case of a quench~\cite{Calabrese2006,Cazalilla2006,Rigol2007,Kollath2007,Manmana2007,Lauchli2008,Moeckel2008,Manmana2009}, the modification is not considered as instantaneous, but rather a finite linear ramp. In addition, 
contrary to the (quasi-)harmonic scenario discussed by Polkovnikov and Gritsev~\cite{Polkovnikov2008} 
we are here particularly interested in the correlation dynamics in the deeply quantum regime 
$U_0\gg J$ at low filling $n_0<1$, where the system remains superfluid although strongly correlated. In this regime, quantum fluctuations are dominant, and hence 
TWA approximation cannot be employed to describe the dynamics. To this aim, we employ 
time-dependent Matrix-Product-state techniques, which allow us to study accurately 
relatively large systems. We show that, as for the case of abrupt quenches 
the correlation dynamics is characterized by an intermediate regime, in which quasi-local 
correlations have converged to a new equilibrium, although long-ranged correlations, and in particular 
the quasi-condensate fraction still presents an observable dynamics (which keeps evolving well after the 
end of the slow ramp). Since the system is not integrable, eventually a new equilibrium is reached. 
We analyze this final state and the adiabaticity of the modification by means of the transferred energy 
and the fidelity with respect to the expected ground state solution.

The scheme of the paper is as follows. Sec.~\ref{sec:Methods} introduces the model under consideration and the numerical methods employed. 
Sec.~\ref{sec:Correlations} is devoted to the analysis of correlation functions and quasi-condensate fraction. Sec.~\ref{sec:Fidelity} 
studies the fidelity of the final evolved state with respect to the ground state of the final Hamiltonian. Finally, Sec.~\ref{sec:Conclusions} summarizes our conclusions.

\section{Model and methodology}
\label{sec:Methods}
In the following we consider bosons in a deep lattice constrained to the lowest energy band. In this regime, the free energy of the system is described by the BHH
\begin{equation}
\hat{H}=\sum_i \left [ 
-J(t)\left(\hat{a}_i^\dagger \hat{a}_{i+1}+h.c\right)+ \frac{U_0}{2}\hat{n}_i\left(\hat{n}_i-1\right)-\mu \hat n_i \right ]
\end{equation}
where $\hat{a}_i$ ($\hat{a}_i^\dagger$) is the annihilation (creation) operator of a boson at the 
$i$-th site, $\hat{n}_i$ is the corresponding number operator, $\mu$ is the chemical potential, $J(t)$ is the time-dependent tunneling amplitude between neighbouring sites, and $U_0 > 0$ is the repulsive on-site interaction. We consider in the following that at $t=0$ the system 
is in the ground-state for the initial $J_i=J(0)$. 

As mentioned above, the system may be driven out of equilibrium either by modifying the hopping rate 
(as we consider here) or by modifying the interaction rate (e.g. by means of Feshbach resonances). 
In the following we consider a time-dependent lattice depth, which leads to a linear-ramp 
$J(t)=J_i+(J_f-J_i)\tfrac{t}{t_r}$ for an initial time interval $0<t<t_r$, 
where $J_f$ is the final hopping. For $t>t_r$ the system evolves 
at a constant $J=J_f$. We consider sufficiently large post-ramp times such that the quantities 
of interest enter into a new equilibrium. 

We consider in our calculations a lattice with $L=60$ sites with open boundary conditions, which is sufficiently large to minimize finite-size effects at the lattice center, where we evaluate the correlations discussed below. In our time-evolution simulations, we work in the canonical ensemble with two different total number of particles, $N=20$ which leads to an average lattice-site filling $\bar{n}\sim 0.3$ below half-filling (HF), and $N=50$ which 
leads to $\bar{n}\sim 0.8$, i.e. above HF. 
In all simulations discussed below we considered 
$J_i=0.1375U_0$ and $J_f=0.2500U_0$. These values were chosen relatively close to each other 
to allow for the convergence to a new equilibrium discussed below within a numerically 
available evolution time. In spite of that these values are sufficiently different to allow for 
the study of the ramping adiabaticity. Note that the hopping rates $J_i$ and $J_f$ are rather low and comparable to the critical tunneling for on-set of MI (tip of the lowest MI lobe), which for 1D 
is found at $J\simeq 0.2U_0$. In that regime quantum fluctuations are highly relevant, but due to the low filling $\bar{n}$ considered, the system remains within the (highly-correlated) SF regime.

In our calculations we first obtained the ground-state at $J=J_i$ by means of an MPS-algorithm 
using a similar approach as that of Refs.~\cite{Verstraete2004a,Verstraete2004b}. In this DMRG-like 
technique, the many-body state is approximated by a MPS ansatz of the form
\begin{equation}
|\Psi_{MPS}\rangle=\sum_{s_1,\dots,s_N=1}^d Tr(A_1^{s_1}\dots A_N^{s_{N}})|s_1,\dots,s_N\rangle,
\end{equation}
where the $A$'s are matrices of a given dimension $D$ and $d$ is associated to the 
on-site dimension, which in our case represents the maximal number of atoms per site considered, (for our parameters it can be safely assumed as $d=2$). The MPS-algorithm performs in an efficient way a variational method using the matrix elements $A_{i}^{s_i}$ as variational parameters, and basically reduces to subsequent DMRG sweeps over the lattice until achieving energy convergence~\cite{Schollwock2005}.

Once the ground state is obtained with $J=J_i$ we evolve the problem in real time for $t>0$ by means of the time-evolving block decimation (TEBD) method~\cite{Vidal2003}. TEBD takes advantage from the fact that the Hamiltonian can be written as the sum over an even and an odd sites and then, the time evolution operator is approximated using the Trotter-Suzuki expansion formula~\cite{Suzuki1990}. Since the entanglement of the system grows on time, to keep the matrices of the same size means to loose information through out the time evolution. In order to avoid that, we adapt the value of $D$ in each step to keep the state as faithful to the real one as possible.


\begin{figure}[t!]
    \includegraphics[scale=0.5]{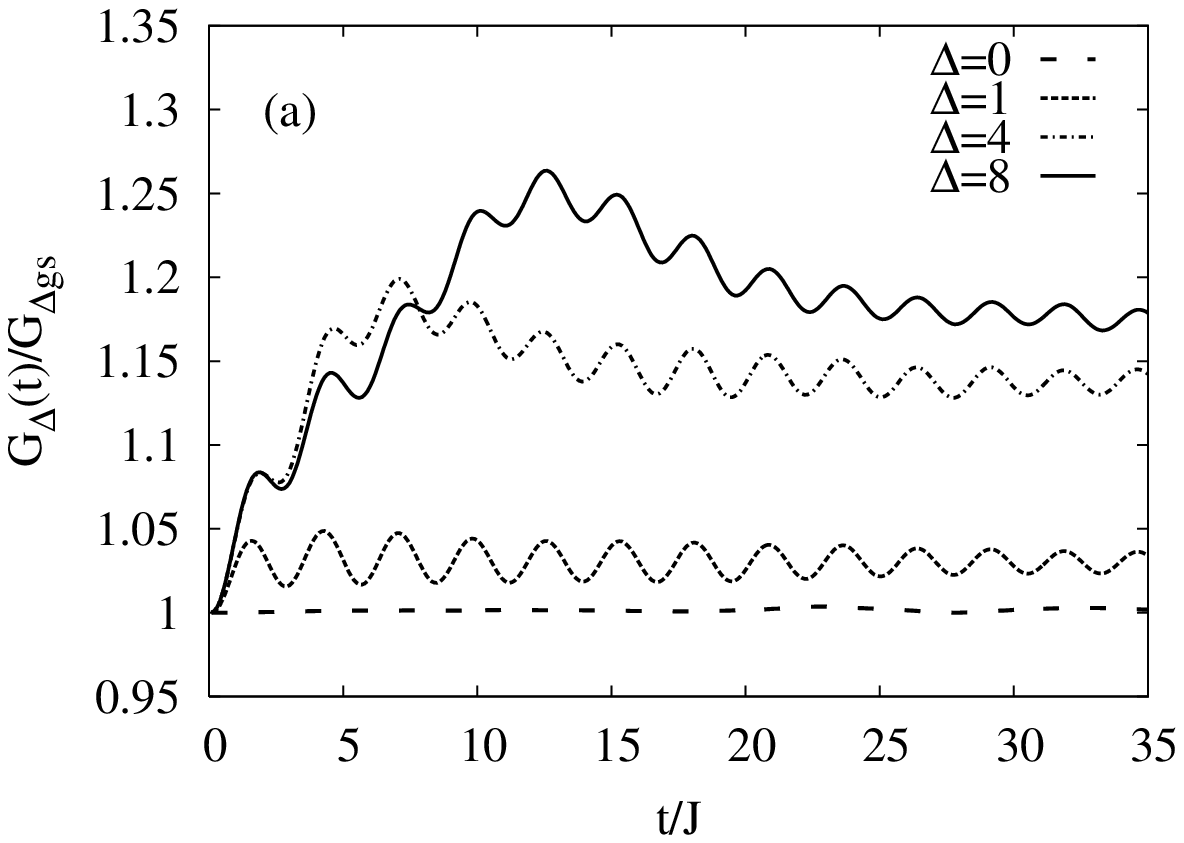}
    \includegraphics[scale=0.5]{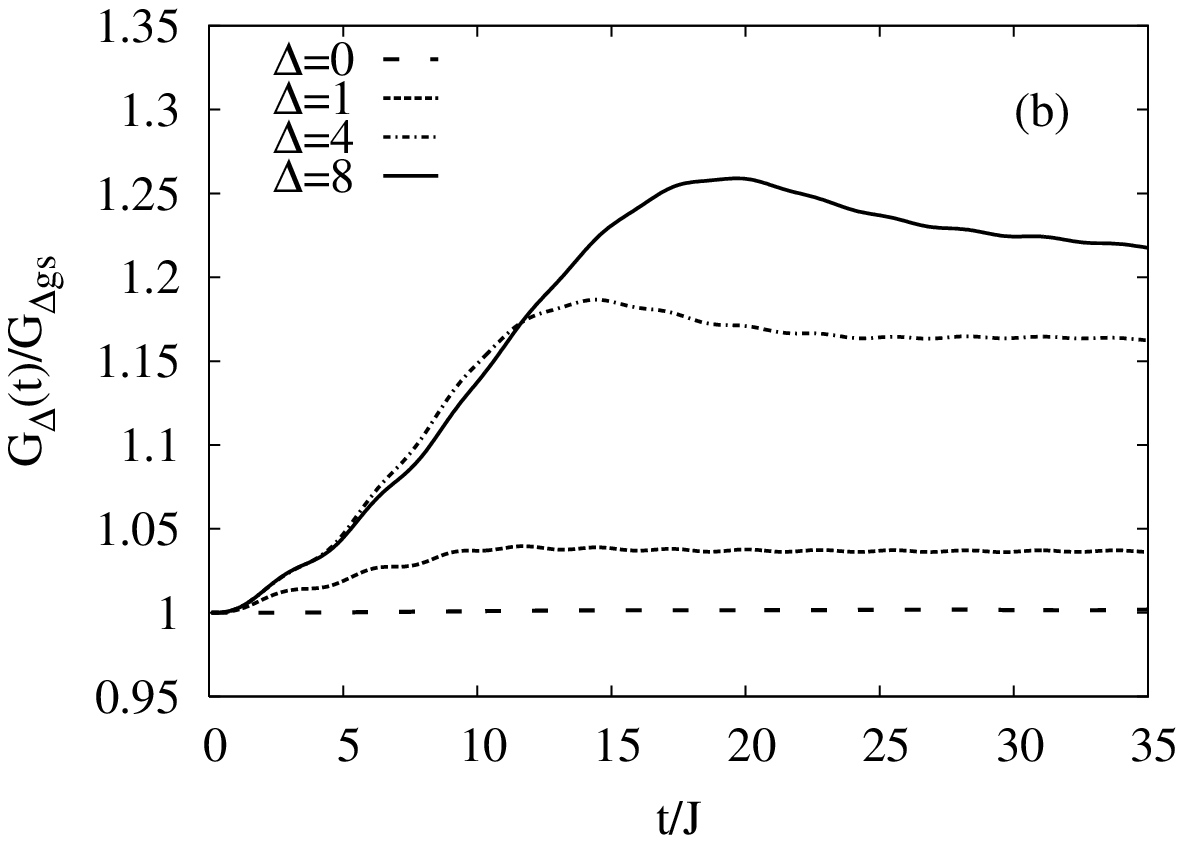}
  \caption{Time-evolution of several  correlation functions considering a ramp-time of (a) $t_r=0.2J_f$ and (b) $t_r=10.0J_f$. The used density population is $\bar n=0.3$.}
\label{fig:1}
\end{figure}


\begin{figure}[t!]
\psfrag{initi}{\hspace{-0.3cm}$|\varphi_{i gs}\rangle$}
\psfrag{final}{\hspace{-0.2cm}$|\varphi_{f gs}\rangle$}
\psfrag{t/J35}{\hspace{-0.2cm}$|\varphi_{evol}\rangle$}
 \includegraphics[scale=0.5]{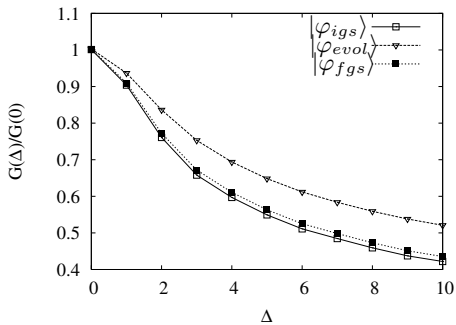}
 \caption{Spatial correlation for the initial state $|\varphi_{i gs}\rangle$, the ground-state $|\varphi_{f gs}\rangle$ of the final configuration and the evolved state $|\varphi_{evol}\rangle$ after $t/J=35$ for a ramp time of $t_r=10.0J_f$.}
\label{fig:2}
\end{figure}
\vspace{-0.5cm}
\section{Correlation dynamics and quasi-condensate fraction}
\label{sec:Correlations}

In the following we are particularly interested on the dynamics of the 
single particle correlation functions 
$G(\Delta;t)\equiv \langle \hat a_j^\dag \hat a_{j+\Delta} \rangle$. In our numerical 
calculations we evaluate $G(\Delta,t)$ at the lattice center $j=0$. In equilibrium, the 
1D SF phase is characterized by a polynomial decrease   
$G(\Delta)\sim \Delta^{-\eta}$ with the relative distance $\Delta$~\cite{Kuhner2000} (contrary to the gapped MI phase where an exponential decay is expected).

We have analyzed the evolution of $G(\Delta,t)/G(\Delta,t=0)$ for different values of $\Delta$ 
and different ramping times $t_r$. Fig.~\ref{fig:1}(a) exemplifies the case of an abrupt 
ramp (basically an instantaneous quench) with $t_r=0.2 J_f$, whereas Fig.~\ref{fig:1}(b) 
depicts typical results observed for a mild ramp, in this case $t_r=10.0 J_f$. Both cases are 
calculated at a filling factor $\bar{n}=0.3$. Note that 
$G(\Delta,t)$ shows in both cases a significant dynamics following the linear ramp (due to number conservation the density, i.e. $\Delta=0$, is unaffected by the lattice modulation). 

Both abrupt and slow ramps lead to an evolution of the correlations characterized by an initial short-time scale, followed by an eventual convergence into a new equilibrium at longer times (observe, however, that the correlation dynamics following the abrupt quench presents a short-time 
modulation which persists well within the quasi-equilibrium region~\cite{Lauchli2008}). The short-time evolution of the correlations continues well after the end ($t=t_r$) of the ramp (even for such a mild ramp). Notice also that short-distance correlations, in particular $\Delta=1$ converge significantly quicker than correlations at larger distances. As a consequence the lattice bosons experience a transient regime characterized by a quasi-equilibrium of local or quasi-local observables coexisting with out-of-equilibrium global properties. After this transient regime the system reaches a final equilibrium, characterized by 
an equilibrium correlation $G(\Delta)$ (Fig.~\ref{fig:2}).

The transient regime becomes particularly clear from an analysis of the quasi-condensate fraction. This fraction 
may be defined as the largest eigenvalue $\lambda_0$ of the density matrix $\langle \hat a_i^\dag \hat a_j \rangle$ of the system~\cite{Legget2001}, and hence may be considered a global property of the system, influenced by correlations at any available $\Delta$. Although strict condensation is prevented in 1D, quasi-condensation, characterized by a distinct finite $\lambda_0$, is possible in finite systems. 

\begin{figure}[t!]
\psfrag{n=20}{\hspace{-0.2cm}\footnotesize{$\bar n=0.3$}}
\psfrag{n=50}{\hspace{-0.2cm}\footnotesize{$\bar n=0.8$}}
\psfrag{th1}{\footnotesize{$t_{1}$}}
\psfrag{th2}{\footnotesize{$t_{2}$}}
\psfrag{y=1.745}{\hspace{-0.7cm}\footnotesize{$\lambda^{qe}=1.75$}}
\psfrag{y=1.220}{\hspace{-0.6cm}\footnotesize{$\lambda^{qe}=1.22$}}
\includegraphics[scale=0.5]{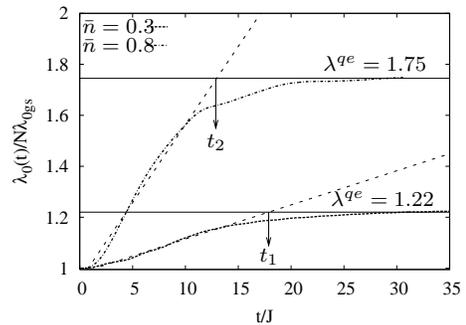}\caption{The evolution of the quasi-condensate fraction for two different particle densities $\bar n=0.3$ and $\bar n=0.8$ in a mild ramp ($t_r=10.0J_f$). The horizontal lines denote where the quasi-equilibrium ($\lambda^{qe}$) values tend in each case. The dashed lines fit the initial curve stretch and the intersection points are defined as a qualitative estimation of the $\lambda_0$ variation time, $t_1(\bar n=0.3)= 17.9$ and $t_2(\bar n=0.8)= 12.9$. The plot is in units of $\lambda_{0 gs}$ where $\lambda_{0 gs}$ is the quasi-condensate fraction of the initial ground-state.}
\label{fig:3}
\end{figure}

The time evolution of $\lambda_0(t)/\lambda_0(0)$ is depicted in Fig.~\ref{fig:3}. Note that the quasi-condensate fraction evolves at a much longer time scale (much larger than the ramp time even in the mild-ramping case) than quasi-local correlations (it has not yet fully converged in our typical calculations). The filling factor ${\bar n}$ is of course important to determine the time scale of variation of the different correlation functions. The figure shows for a ramp $t_r=10 J_f$ the 
evolution of the quasi-condensate fraction for the  
case of an average filling factor $\bar{n}=0.8$ compared to the case with $\bar{n}=0.3$. 
At higher densities the quasi-equilibrium is reached faster for the same ramping (a similar behaviour is observed for $G(\Delta)$). As in Fig.~\ref{fig:3} we may define a typical time scale for 
the variation of $\lambda_0$, which is $t ({\bar n}=0.3)=17.9$ and $t({\bar n}=0.8)=12.9$.

\vspace{-0.5cm}
\section{Fidelity and final energy}
\label{sec:Fidelity}

\begin{figure}[t!]
\psfrag{Fidelity}{\hspace{-0.3cm}\small{Fidelity ($\mathbb{F}$)}}
\psfrag{t0.2}{\hspace{-1.1cm}\small{$t_r/J_f=0.2$}}
\psfrag{t10.}{\hspace{-1.1cm}\small{$t_r/J_f=10.$}}
\psfrag{t20.}{\hspace{-1.1cm}\small{$t_r/J_f=20.$}}
\psfrag{t30.}{\hspace{-1.1cm}\small{$t_r/J_f=30.$}}
\includegraphics[scale=0.6]{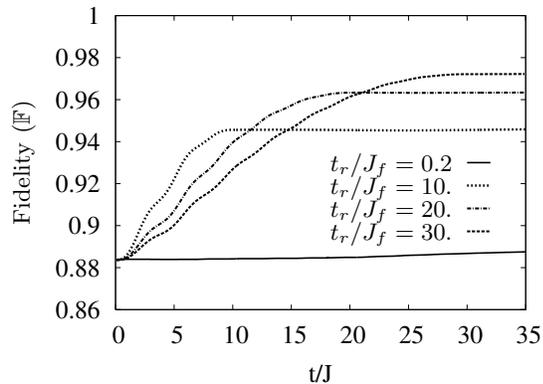}
\caption{Fidelity of the time-evolved state to the ground state at the end of the lattice modulation for $t_r/J_f=0.2$,  $t_r/J_f=10.0$, $t_r/J_f=20.0$ and $t_r/J_f=30.0$ time modulation.}
\label{fig:4}
\end{figure}

A good tool for the analysis of the adiabaticity of the ramping is provided 
by the fidelity $\mathbb{F}=\left |\langle \varphi_{fgs}|\varphi(t) \rangle\right |^2$, 
i.e. the Hilbert-space distance between the time-evolved state 
$ |\varphi(t) \rangle$ and the expected ground-state $|\varphi_{fgs}\rangle$ 
calculated with the final $J=J_f$. The fidelity $\mathbb{F}$ is an interesting figure of merit for the 
adiabaticity of the ramping, since contrary to other figures (as the correlation functions discussed above) it just evolves while the ramping is on, since after the ramping the eigenstates of the final 
Hamiltonian are of course stationary, and consequently $\mathbb{F}$ remains constant. Hence, although other quantities require a rather long waiting time for comparing the final state and the time-evolved one, 
$\mathbb{F}$ provides an answer at the relatively short-time scale $t_r$ (see Fig.~\ref{fig:4}). As expected the abrupt ramping $t_r/J_f=0.2$ leads basically to an instantaneous projection of the initial ground-state $|\varphi_{fis}\rangle$
into $|\varphi_{fgs}\rangle$ (note that the overlapping is already rather large, $88\%$, due to the 
relative close values of $J_i$ and $J_f$). As expected the milder ramping approaches further to $|\varphi_{fgs}\rangle$, however, the fidelity is still $5\%$ off from $|\varphi_{igs}\rangle$. Interestingly, this indicates 
that even very large ramping times significantly larger than the hopping time, and for a relatively small variation of $\Delta J=0.1125U_0$, do not guarantee a perfect transfer into the ground state of the final configuration. The analysis of $\mathbb{F}$ for even milder ramps shows (see Fig.~\ref{fig:4}) that milder ramps lead indeed to more adiabatic transfers (contrary to what may be expected in the harmonic regime~\cite{Polkovnikov2008}). 

Hence, although, as mentioned above, 
the correlations $G(\Delta)$ approach at longer times to a new equilibrium, 
this new equilibrium is not that given by the expected ground state, but by a new 
distribution with a higher energy. The analysis of the final energy after the ramping does 
not provide however an equally strict adiabaticity analysis as that of the fidelity, especially 
for situations as those discussed here, in which $J_i$ and $J_f$ possess relatively close values. 
In our case the difference between the energy of $|\varphi_{igs}\rangle$ (or equivalently that of the system after the abrupt ramping) and 
that of $|\varphi_{fgs}\rangle$ is less than $1\%$, whereas the time-evolution with the mild ramping 
provides a final energy less than $0.1\%$ off that of $|\varphi_{fgs}\rangle$.

\vspace{-0.4cm}
\section{Conclusions}
\label{sec:Conclusions}

We have analyzed by means of MPS techniques the dynamics of correlation functions 
and quasi-condensate fraction of ultra cold lattice bosons in the deeply correlated superfluid regime
during and after a finite linear ramp modulation of the hopping rate. We have shown that the evolution is characterized by a transient non-equilibrium state in which quasi-local correlation functions have already converged into a new equilibrium
whereas long-range correlations and the quasi-condensate fraction present still a significant 
time dependence. Additionally, we have analyzed the formation (at a longer time scale) 
of a new equilibrium from an initial gas at zero temperature. By considering the fidelity 
with respect to the ground state of the final configuration we have shown that even 
rather mild ramps do not fully guarantee a perfect loading of the new ground state. We have however shown that contrary to the harmonic regime~\cite{Polkovnikov2008} progressively milder ramps lead to a more adiabatic transfers.
\vspace{-0.6cm}
\begin{acknowledgements}
 We thank A. Polkovnikov for his interesting remarks. This work is supported by the European Graduate College ``Interference and Quantum Applications'' and the Deutsche Forschungsgemeinschaft (QUEST,SFB 407).
\end{acknowledgements}


\begin{thebibliography}{99}

\bibitem{Greiner2002} M. Greiner, O. Mandel, T. Esslinger, T. W. H¨ansch, and
I. Bloch, Nature {\bf 415}, 39 (2002).
\bibitem{Jordens2008}  R. J\"ordens, N. Strohmaier, K. G\"unter, H. Moritz, and T.
Esslinger, Nature {\bf 455}, 204 (2008).
\bibitem{Schneider2008} U. Schneider, L.Hackerm\"uller, S. Will, Th.Best, I. Bloch,
T. A. Costi, R. W. Helmes, D. Rasch, and A. Rosch, Science {\bf 322}, 1520 (2008).
\bibitem{Paredes2004} B. Paredes, A. Widera, V. Murg, 
O. M. S. F\"olling, I. Cirac, G. V. Shlyapnikov, T. W. H\"ansch, and I. Bloch,
Nature {\bf 429}, 277 (2004).

\bibitem{Calabrese2006} P. Calabrese and J. Cardy, Phys. Rev. lett. {\bf 96} 136801 (2006). 
\bibitem{Cazalilla2006} M. Cazalilla, Phys. Rev. Lett. {\bf 97}, 156403 (2006).
\bibitem{Rigol2007} M. Rigol, V. Dunjko, V. Yurovsky, and M. Olshanii, Phys. Rev. Lett. {\bf 98}, 050405 (2007)
\bibitem{Kollath2007} C. Kollath, A. M. L`\"auchli, and E. Altman, Phys. Rev. Lett. {\bf 98}, 180601 (2007).
\bibitem{Manmana2007} S. R. Manmana, S. Wessel, R. M. Noack, and A. Muramatsu,
Phys. Rev. Lett. {\bf 98}, 210405 (2007).
\bibitem{Lauchli2008} A. M. L`\"auchli and C. Kollath, J. Stat. Mech.: Theory Exp. (2008) P05018.
\bibitem{Moeckel2008} M. Moeckel and S. Kehrein, Phys. Rev. Lett. 100, 175702 (2008)
\bibitem{Manmana2009} S. R. Manmana, S. Wessel, R. M. Noack, and A. Muramatsu,
Phys. Rev. B {\bf 79}, 155104 (2009).
\bibitem{Kinoshita2006} T. Kinoshita, T. Wenger, and D. Weiss, Nature (London) {\bf 440}, 900 (2006).

\bibitem{Blakie2004} P. B. Blackie and J. V. Porto, Phys. Rev. A {\bf 69}, 013603 (2004).
\bibitem{Rey2006} A. M. Rey, G. Pupillo, and J. V. Porto, Phys. Rev. A {\bf 73}, 023608 (2006).
\bibitem{Pollet2008} L. Pollet, C. Kollath, K. van Houcke, and M. Troyer, New J. Phys. {\bf 10}, 065001 (2008).

\bibitem{Polkovnikov2008} A. Polkovnikov and V. Gritsev, Nat. Physics {\bf 4}, 477 (2008).

\bibitem{Verstraete2004a} F. Verstraete, D. Porras, and J. I. Cirac, Phys. Rev. Lett.
{\bf 93}, 227205 (2004).

\bibitem{Verstraete2004b} F. Verstraete, J. Garcia-Ripoll, and J. I. Cirac, Phys.
Rev. Lett. {\bf 93}, 207204 (2004).

\bibitem{Schollwock2005} U. Schollw¨ock, Rev. Mod. Phys. {\bf 77}, 259 (2005).

\bibitem{Vidal2003} G. Vidal, Phys. Rev. Lett. {\bf 91}, 147902 (2003).

\bibitem{Suzuki1990} M. Suzuki, Phys. Lett. A {\bf 146}, 319 (1990).

\bibitem{Kuhner2000} T. K\"uhner, W. White, and H. Monien, Phys. Rev. B {\bf 61},
12474 (2000).

\bibitem{Legget2001} A. J. Leggett, Rev. Mod. Phys. {\bf 73}, 307 (2001).




\end{thebibliography}
\end{document}